\pdfoutput=1

\documentclass[conference]{IEEEtran}
\ifCLASSINFOpdf
\else
\fi

\usepackage{pgf}
\usepackage{tikz}
\usetikzlibrary{arrows,automata}
\usepackage{inputenc}

\usepackage{tabularx,booktabs}
\usepackage{multirow}

\begin{document}
%
\title{An Empirical Study of Affiliate Marketing Disclosures on YouTube and Pinterest}

\author{\IEEEauthorblockN{Arunesh Mathur, Arvind Narayanan, Marshini Chetty\\Princeton University\\\{amathur, arvindn, marshini\}@cs.princeton.edu}}


%


\maketitle

\begin{abstract}
While disclosures relating to various forms of Internet advertising are well established and follow specific formats, endorsement marketing disclosures are often open-ended in nature and written by individual publishers. Because such marketing often appears as part of publishers' actual content, ensuring that it is adequately disclosed is critical so that end-users can identify it as such. In this paper, we characterize disclosures relating to affiliate marketing---a type of endorsement based marketing---on two popular social media platforms: YouTube \& Pinterest. We find that only roughly one-tenth of affiliate content on both platforms contains disclosures. Based on our findings, we make policy recommendations geared towards various stakeholders in the affiliate marketing industry, highlighting how both social media platforms and affiliate companies can enable better disclosure practices.
\end{abstract}


%
\IEEEpeerreviewmaketitle

\section{Introduction}
Advertising on the Internet appears in various forms: as embedded content on websites and emails (e.g. Google AdSense served on The New York Times’ website), as native advertising on websites (e.g. Twitter's \emph{promoted} tweets), and as products endorsed by individual users---often termed \emph{content publishers} or \emph{influencers}---on user-generated content websites such as social media and blogs.

To ensure that users can identify advertisements as such, several regulations require advertisers to place disclosures on or around the advertisements. For instance, AdChoices \cite{adchoices} is a self-regulatory program followed by advertisers in the United States (US), Canada and Europe, which requires them to inform users that data about them and their activities is used to serve tailored advertisements. Similarly, social networks such like Facebook and Twitter tag native advertisements as \emph{sponsored} and \emph{promoted} respectively to distinguish them from regular user-generated content.

While both these kinds of disclosures are largely standardized by each advertising firm or body---whether that be that in the form of text or icons---endorsement advertising disclosures by individual content publishers are largely un-standardized and open-ended in nature. In the United States, these disclosures are guided by the Federal Trade Commission (FTC), which requires that publishers clearly and unambiguously disclose their relationship with merchants and brands. In recent times, the FTC has lodged several cases against publishers who have failed to adequately disclose their relationships with brands and merchants~\cite{ftc1,ftc2,ftc3}. 

However, while such regulatory authorities provide guidelines for publishers to follow, it is unclear how many do, and for the ones that do disclose, whether they follow the guidelines put forward. Ensuring that publishers follow these disclosure guidelines is particularly important, since content that they advertise often appears together with their original content, and may therefore be harder for users to identify.

In this paper, we specifically examine disclosures accompanying \emph{affiliate marketing}---an endorsement marketing strategy that pays affiliates (the content publishers) money when users click on their customized URLs---on YouTube and Pinterest, two popular social media platforms. We investigate affiliate marketing's growth over the years on these platforms, the kind of affiliates that engage in affiliate marketing, the companies that enable affiliate marketing, and finally, how affiliates disclose their relationship with these companies to end-users.

We have three main findings. First, we find that---on both YouTube and Pinterest---content with affiliate URLs have significantly higher user engagement compared to content that does not contain affiliate URLs. Second, we find that affiliate marketing disclosures appear in three distinct formats---which we term \emph{Affiliate Link} disclosures, \emph{Explanation} disclosures, and \emph{Support Channel} disclosures---as opposed to one standardized format. Third, we find that the overall prevalence of these disclosures is low: only $\sim$10\% of all affiliate content on each platform has accompanying disclosures. 

Based on our findings, we analyze how the disclosures we discovered comply with existing FTC disclosure guidelines. We forge policy recommendations for social media platforms and affiliate marketing firms to consider in order to enable affiliates to disclose such marketing relationships easily and clearly. We also outline directions for future work.

\section{Affiliate Marketing}
\begin{figure}
\centering

\begin{tikzpicture}
\node[draw,circle] (A) at (90:2) {Affiliate};
\node[draw,circle] (B) at (210:2) {Merchant};
\node[draw,circle] (C) at (330:2) {Network};
\draw[-latex] (A) to[bend right=10] node[above,rotate=60] {2} (B);
\draw[-latex] (A) to[bend right=10] node[below,rotate=300] {1} (C);
\draw[-latex] (C) to[bend right=10] node[above,rotate=300] {3} (A);
\draw[-latex] (B) to[bend right=10] node[below] {3} (C);
\draw[-latex] (B) to[bend left=10] node[below] {1} (C);
\end{tikzpicture}

  \caption{An Overview Of Affiliate Marketing. (1) Affiliates and Merchants Register With an Affiliate Network. (2) Affiliates Drive Traffic to Merchants through the Affiliate Network. (3) Merchants Pay the Affiliate Network For Each Sale, Who In Turn Pay the Affiliate.}
\label{fig:aff_mark}
\end{figure}
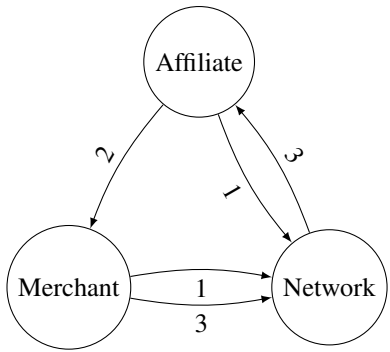

Affiliate marketing primarily involves three entities: the affiliate (sometimes called publisher), the merchant and the affiliate network, and consists of an agreement between the affiliate and the affiliate network, and the affiliate network and the merchant. As illustrated in Figure~\ref{fig:aff_mark}, affiliate marketing typically works as follows: 

\begin{enumerate}
    \item Merchants, affiliates sign up with an affiliate network
    \item Affiliates associate with merchants they wish to promote through the affiliate network
    \item Affiliate networks publish customized URLs for affiliates to distribute along with their content
    \item Every time a sale is made through the URLs, the merchant pays the affiliate a cut of the sale through the affiliate network
\end{enumerate}

In the United States, the FTC's endorsement guidelines \cite{ftc_guidelines} have largely described the disclosure standards for affiliates to maintain. The FTC states that \emph{if there exists a connection between an endorser and the marketer that consumers would not expect and it would affect how consumers evaluate the endorsement, that connection should be disclosed}. Specifically towards affiliate marketing, the guidelines state that disclosures need to be placed close to the recommendation, and not buried inside an \emph{About Us} or \emph{Terms of Service} pages.

While previous research has examined fraud in affiliate marketing \cite{chachra2015affiliate}, and how affiliate programs drive spam revenue and sale of unauthorized items \cite{mccoy2012,kanich2011show}, we know very little about affiliate marketing disclosures. In this work, we examine the prevalence of affiliate marketing disclosures, and whether they are in compliance with the FTC's guidelines.

\section{Method}
In this section, we describe our methodology: our data collection process, and the techniques we employed to identify the affiliate URLs and disclosures.

\subsection{Data Collection}
We examined affiliate marketing on YouTube and Pinterest, two social media platforms that are designed to share reviews and content. To ensure that our investigation uncovered the most accurate state of affiliate marketing and its disclosures on these platforms, we wanted to gather the least possibly biased sample of videos and pins. One strategy to collect videos and pins is to sample from the \emph{related} videos and pins graphs on the network. Sampling from such large graphs has been extensively studied in the literature \cite{leskovec2006sampling}, with each approach having its benefits and drawbacks. For instance, random walks have been shown to produce less biased samples compared to breadth-first and depth-first searches. 

However, because \emph{related} graphs have non-randomly selected edges which are often biased towards content with high engagement---such as videos with higher view counts in the case of YouTube~\cite{Zhou:2011:CYV:2068816.2068851}---they result in non-uniform samples. Another strategy is using keyword searches, where videos are sampled by selecting results arising from specific search terms. However, such samples are also likely to be biased towards the search terms and the platforms' search algorithms.

To collect a more representative and less biased sample, we employed prefix sampling which has previously been used for sampling from YouTube \cite{Zhou:2011:CYV:2068816.2068851} and Pinterest \cite{Gilbert:2013:INT:2470654.2481336}. Prefix sampling works by gathering, ahead of time, part of the identifier of each record called the \emph{prefix}, which is then used to sample records beginning with that prefix. If the prefixes are uniformly gathered or generated, then the resulting samples will be uniform too. This sampling methodology is particularly useful when the search space of identifiers is significantly larger than the number of already issued identifiers, and it is not possible to generate issued identifiers randomly.

On YouTube, videos are assigned an eleven character long identifier. In order to gather a uniform sample, we first randomly generated video identifiers of length five. We then searched for those five character long identifiers using the YouTube Search API \cite{YouTube_API}. This, by way of a feature of the API, returned a list of videos beginning with that prefix. Similarly on Pinterest, pins are assigned a varying length identifier consisting wholly of numbers. The last five digits of the identifier however, represent a timestamp. To gather the prefixes, we retrieved twenty five pins from the Pinterest \emph{Categories} page \cite{Pinterest_Categories}, and sampled their related pins randomly. We then varied the last five digits from zero to ten thousand to retrieve the sample of pins. In total, we retrieved 515,999 unique YouTube videos and 2,140,462 unique Pinterest pins. While retrieving the videos and pins, we also recorded their characteristics such as their categories, view counts, comment counts, and details about their creators. We collected this data between August and September 2017.

\subsection{Data Analysis}

\subsubsection{Discovering Affiliate URLs}
After sampling the videos and pins, we began identifying affiliate URLs from our corpus. First, we gathered a list of all URLs from the descriptions of the YouTube videos and the URLs of the Pinterest pins. Next, we resolved each URL following both server-side and client-side redirects (HTTP 3XX, Meta refresh), and recorded the resulting URLs and their HTTP response codes.

To identify affiliate URLs from this set of resolved URLs, we relied on the observation that affiliate URLs contain predictable patterns. For instance, Amazon's affiliate URL contains a \emph{tag} parameter which indicates the identifier of the affiliate who stands to gain money from the purchase. While Amazon's affiliate URL appears at the end of a redirect---to Amazon's website---affiliate URLs may also appear during the intermediate redirects. Further, unlike Amazon, affiliate URL patterns may not necessarily only emerge as URL parameters; they may also be present in other parts of the URL including its path and sub-domain.

To ensure that our analysis discovered all such cases, we performed a frequency analysis using each resolved URL's domain, sub-domain, path, and parameters, creating a list of commonly occurring patterns sorted by decreasing order of appearance (counts). We reasoned that if there existed any patterns across the URLs we resolved and visited, our frequency analysis would capture and bubble those to the top of the list. Starting with the sub-domain and path, we first recorded how many sub-domains (paths) each domain appeared with. A high number of sub-domains (paths) would signal that an affiliate marketing company likely caters to different merchants through unique sub-domains (paths), or that a constant sub-domain (path) appears as part of the affiliate URL. Next, we turned our attention to domains and their URL parameters. Rather than recording the number of parameters associated with each domain, we recorded the number of times a domain appeared with a URL parameter. A frequent co-occurrence of domains and URL parameters would signal that the parameter conveys some information about the merchant/affiliate to the affiliate marketing company. 

Because these lists contained a high number of false positives, we manually scanned each list, examining which of the domains, sub-domains, paths, parameters corresponded to affiliate marketing companies. To limit the effort required to manually examine these lists, we only examined those combinations of domains and sub-domains/paths/parameters that appeared at least 15 times. We also queried the FMTC affiliate database \cite{FMTC}, and where possible, signed up on these programs as affiliates to validate our findings.

\subsubsection{Characterizing Disclosures}
Once we finalized the affiliate URL patterns, we first filtered the list of resolved URLs to only retain those corresponding to these patterns. We then filtered the YouTube videos and Pinterest pins datasets to only those containing the affiliate URLs. To extract the disclosures present in these videos' and pins' descriptions, we first split each description by its newlines and then by the sentences contained in each newline. We then tokenized each resulting sentence into a bag-of-words representation, and clustered the sentences using hierarchical clustering \cite{Rokach2005} with the euclidean distance metric. We chose a fairly low cut-off for the clusters based on the idea that the relevant smaller clusters containing the affiliate disclosures may already have been formed at that cut-off. We then manually examined these clusters one after the other, and recorded ones that contained disclosures pertaining to affiliate marketing. For this analysis, we only considered those descriptions that were written in English.

\section{Findings}
Across YouTube, we discovered a total of 0.67\% or 3,472 of 515,999 videos, and across Pinterest, a total 0.85\% or 18,237 of 2,140,462 pins contained at least one affiliate URL. In this section, we present our findings, describing the characteristics of affiliate content on both platforms, the types of disclosures we discovered, and the characteristics of these disclosures. 

\subsection{Affiliate Marketing Companies}

Table~\ref{fig:affiliate} in the Appendix lists the affiliate marketing companies we discovered along with their URL patterns, and the number of times we observed their presence across all URL resolutions. Within each platform, Amazon's Amazon Associate Program\footnote{https://affiliate-program.amazon.com/} had the largest presence (YouTube = 7,308, Pinterest = 7,368), closely followed by AliExpress' Affiliate Program\footnote{https://portals.aliexpress.com} (YouTube = 2,167, Pinterest = 785). We also discovered certain merchants hosted in-house affiliate programs, as opposed to explicitly redirecting through an affiliate marketing company. For instance, Booking.com\footnote{https://www.booking.com/affiliate-program/v2/index.html} and Apple\footnote{https://www.apple.com/itunes/affiliates/} marketed products through their own affiliate programs.

\subsection{Characteristics of Affiliate Content}

\begin{table}[t]
  \centering
  \caption{Percentage of Affiliate Content by Category on YouTube and Pinterest. Only Top 10 Categories are Listed.}
\label{fig:category}
\begin{tabular}{l c | l c}
\toprule
\textbf{YouTube Category} & \textbf{Perc.} & \textbf{Pinterest Category} & \textbf{Perc.}\\
\midrule

Science \& Technology & 3.61  & Women's Fashion & 4.62  \\

Howto \& Style & 3.49  & Products & 2.21  \\

Travel \& Events & 1.93  & Hair \& Beauty & 2.04  \\

Film \& Animation & 1.59   & Sports & 1.63  \\

Shows & 1.36  & Design & 1.53  \\

Music & 0.94  & Outdoors & 1.21  \\

Entertainment & 0.68  & Technology & 1.16  \\

Education & 0.64  & Men's Fashion & 1.12  \\

Gaming & 0.63  & Animals & 1.00  \\

People \& Blogs & 0.39 & \emph{No Category} & 0.92  \\

\bottomrule
\end{tabular}
\end{table}

\subsubsection{Affiliate Content Categories}

We found that across YouTube and Pinterest there existed several similarities in the categories with highest affiliate content. Overall, affiliate content on both platforms was dominated by affiliates working with fashion, beauty, and style merchants. Prevalence across YouTube's \emph{Howto \& Style} category stood at 3.49\%, and Pinterest's \emph{Women's Fashion} and \emph{Hair \& Beauty} categories stood at 4.62\% and 2.04\% respectively. Similar patterns existed in the travel category, with YouTube's outdoors \emph{Travel \& Events} at 1.93\%, and Pinterest's \emph{Outdoors} at 1.21\%. 

There also existed dissimilarities. For instance on YouTube, affiliate content was most popular in the \emph{Science \& Technology} category (3.61\%) whereas Pinterest's \emph{Technology} category stood at 1.16\%. The list of categories ordered by their affiliate content is presented in Table~\ref{fig:category}.

\subsubsection{Affiliate Content Engagement Metrics}

In addition to examining the prevalence of affiliate content by content category, we examined how affiliate and non-affiliate content (content without affiliate URLs) correlated with engagement metrics such as view, like, and comment counts. We conducted Mann--Whitney \emph{U} tests to assess statistical significance, correcting for multiple testing using the Bonferroni method.

Across both YouTube and Pinterest, we noted a common thread: affiliate content correlated with higher engagement metrics. On YouTube, affiliate content was longer in duration ($U \sim 7.95 \times 10^8$, $p < 0.0001$), had higher view counts ($U \sim 7.72 \times 10^8$, $p < 0.0001$), had higher like counts ($U \sim 6.96 \times 10^8$, $p < 0.0001$), and had higher dislike counts ($U \sim 6.52 \times 10^8$, $p < 0.0001$). Similarly, on Pinterest, affiliate content had higher repin counts ($U \sim 1.93 \times 10^{10}$, $p < 0.0001$). We could not directly compare the like and comment counts on Pinterest since the Pinterest API returned both as zero for the pins in our dataset.

\subsection{Types of Affiliate Disclosures}

Across YouTube and Pinterest, we discovered that 10.49\% of all affiliate videos and 7.03\% of all affiliate pins contained an affiliate marketing disclosure . In total, we discovered three distinct types of disclosures; Table~\ref{fig:prevdisc} summarizes our findings. 


\subsubsection{``Affiliate Link'' Disclosures}

The first type of affiliate disclosure we discovered communicated to users that affiliate URLs were present in the content. On YouTube, these disclosures appeared in the video description either as blanket disclosures---a single disclosure across the entire description---or as disclosures highlighting the individual affiliate URLs, or both. Disclosures of this type were present in 7.02\% of all affiliate videos. The following statements describe them:

\begin{itemize}
\item \emph{Affiliate links may be present above}
\item \emph{Some of the links may be affiliate links}
\item \emph{(Disclosure: These are affiliate links)}
\item \emph{*Amazon link(s) are affiliate links}
\end{itemize}

On Pinterest, these disclosures appeared in similar formats in the description of the pins, and were present on 4.60\% of all affiliate pins. Unlike YouTube, these disclosures did not point to specific URLs, since the pins only contained one URL: the URL actually pinned. The following statements describe them:

\begin{itemize}
    \item \emph{(aff link)}
    \item \emph{(affiliate)}
    \item \emph{\#affiliatelink}
    \item \emph{This is an Amazon Affiliate link}
\end{itemize}

\subsubsection{Explanation Disclosures}

The second kind of affiliate disclosure we discovered offered users a verbose explanation about affiliate marketing and affiliate URLs, and how clicking on affiliate URLs impacts users. Relative to \emph{Affiliate link} disclosures, these types of disclosures were more detailed, and often quoted specific merchants or affiliate marketing companies. On YouTube, these disclosures were present in 1.82\% of all affiliate videos. The following statements describe them:

\begin{itemize}
    \item \emph{This video contains affiliate links, which means that if you click on one of the product links, I'll receive a small commission}
    \item \emph{I am an affiliate with eBay, Amazon, B\&H and Adorama, which means I get a small commission when you buy through my links}
    \item \emph{**Links that start with http://rstyle, Beautylish \& MUG links are affiliate links, I do earn a small commission when you purchase through them, which helps me purchase products for review \& improve my channel}
\end{itemize}

On Pinterest, these disclosures appeared in similar formats in the description of the pins, and were present in 2.43\% of all affiliate pins. The following statement describe them:

\begin{itemize}
    \item \emph{(This is an affiliate link and I receive a commission for the sales)}
\end{itemize}

\subsubsection{Channel Support Disclosures}

The third kind of affiliate disclosures we discovered communicated to users they would be supporting the channel by clicking on the affiliate URLs, without explaining how exactly. These disclosures appeared exclusively on YouTube, and were present in 2.44\% of all affiliate videos. The following statements describe them:

\begin{itemize}
    \item \emph{AMAZON LINK: (Bookmark this link to support the show for free!!!)}
    \item \emph{Support HWC while shopping at NCIX and Amazon}
    \item \emph{Purchase RP here and help support this channel via the amazon affiliate program (NA): http://amzn.to/}
\end{itemize}

\begin{table}[t]
  \centering
  \caption{Prevalence of Affiliate Disclosures On YouTube and Pinterest.}
\label{fig:prevdisc}
\begin{tabular}{l c c}
\toprule
\textbf{Disclosure Type} & \textbf{YouTube Perc.} & \textbf{Pinterest Perc.}\\
\midrule

\emph{Affiliate Link} Disclosures & 7.02 & 4.60 \\
\emph{Explanation} Disclosures & 1.82 & 2.43 \\
\emph{Channel Support} Disclosures & 2.44 & NA \\

\bottomrule
\end{tabular}
\end{table}

\subsection{Disclosure Characteristics}

\subsubsection{Prevalence by Content Category}

We examined how the disclosures varied by content category on both YouTube and Pinterest to investigate whether certain categories of content---and therefore, merchants and affiliate marketing companies---are more or less likely to disclose than others. Rather than examining the prevalence of each type of disclosure, we considered all content that had at least one of the three types of disclosure. We also limited our analysis to those content categories for which we had at least 100 samples in our dataset. Table~\ref{fig:prev_category_youtube} and Table~\ref{fig:prev_category_pinterest} summarize our findings.

\begin{table}[t]
  \centering
  \caption{Prevalence of Disclosures by Category on YouTube. Only Categories with more than 100 Samples are Listed. Only the top 10 categories are listed.}
\label{fig:prev_category_youtube}
\begin{tabular}{l r r}
\toprule
\textbf{YouTube Category} & \textbf{Prevalence (Perc.)} & \textbf{Affiliates Disclosing} \\
\midrule

Howto \& Style & 23.68 & 59 of 246 \\

Gaming & 11.22 & 21 of 152\\

Entertainment & 11.21 & 21 of 207  \\

Science \& Technology & 9.72  & 13 of 133 \\

People \& Blogs & 8.87 & 40 of 443   \\

Film \& Animation & 4.86  & 9 of 143 \\

Travel \& Events & 4.12 & 7 of 150 \\

Music & 1.77 & 6 of 366 \\

\bottomrule
\end{tabular}
\end{table}

\begin{table}[t]
  \centering
  \caption{Prevalence of Disclosures by Category on Pinterest. Only Categories with more than 100 Samples are Listed. Only the top 10 categories are listed.}
\label{fig:prev_category_pinterest}
\begin{tabular}{l r r}
\toprule
\textbf{Pinterest Category} & \textbf{Prevalence (Perc.)} & \textbf{Affiliates Disclosing} \\
\midrule
Animals & 60.67 & 26 of 71 \\
Food \& Drink & 14.08 & 7 of 68 \\
Health \& Fitness & 12.76 & 5 of 45 \\
Hair \& Beauty & 11.37 & 20 of 238 \\
Kids & 8.33 & 6 of 74 \\
Home Decor & 7.32 & 18 of 248 \\
DIY Crafts & 5.23 & 14 of 162 \\
Women's Fashion & 4.21 & 39 of 960 \\
\bottomrule
\end{tabular}
\end{table}

Overall, we found that the coverage of affiliate disclosures across content categories varied significantly, with certain categories like Pinterest's \emph{Animals} as high as 60\%, and others such as YouTube's \emph{Music} as low as 1.77\%. On YouTube, the highest disclosure coverage occurred in the \emph{Howto \& Style} category with 23.68\%.

\subsubsection{Prevalence by Year}

In addition to examining the distribution of affiliate disclosures by content category, we also investigated how the affiliate disclosure coverage varied over the past years. Again as before, we only restricted our analyses to those years---across both YouTube and Pinterest---that contained a minimum of 100 samples. We examined the overall rate of disclosures and did not control for possible confounders such as the types of disclosure, the category of content, or the affiliate's country since the prevalence of disclosures was low. Table~\ref{fig:prev_year} lists our findings.

\begin{table}[t]
  \centering
  \caption{Prevalence of Disclosures by Year on YouTube and Pinterest.}
\label{fig:prev_year}
\begin{tabular}{l r r}
\toprule
\textbf{Year} & \textbf{YouTube Perc.} & \textbf{Pinterest Perc.}\\
\midrule

2013 & 9.80  & 0.00 \\
2014 & 8.74  & 0.00  \\
2015 & 12.61  & 0.00  \\
2016 & 12.36  & 1.93  \\
2017 & 8.99  & 11.20  \\

\bottomrule
\end{tabular}
\end{table}

Overall, across both platforms we found that the rate of disclosures by year remained mostly steady. Unlike Pinterest, where we noticed no disclosure coverage until 2016, YouTube had disclosures first appear in 2013 in our dataset.

\section{Discussion}
In this section, we discuss the broader implications of our findings, and highlight directions for future work.

\subsection{Have the FTC's Endorsement Guidelines Been Effective?}

In the United States, the FTC first updated its affiliate marketing guidelines in 2013, stating that affiliate links embedded inside reviews must be disclosed to end-users so they can decide how much weight to provide to the publisher's endorsement. In fact, in its current version of the guidelines, the FTC highlights that simply stating \emph{affiliate link} is not enough, as users may not understand what that means. Instead, the FTC recommends using a short phrase such as \emph{I get commissions for purchases made through links in this post}---similar to the \emph{Explanation} disclosure---close to the recommendation.

Concerningly, our results show that the overall prevalence of affiliate disclosures is low, and that the disclosures are largely of the variety the FTC specifically advocates against: the \emph{Affiliate Link} disclosures. In fact, \emph{Explanation} disclosures---which the FTC recommends---only appear in 1.82\% and 2.43\% of affiliate content on YouTube and Pinterest respectively.

This presents an opportunity for further examination: why is it that despite the FTC's request for specific disclosure requests in affiliate marketing, affiliates fail to follow the guidelines? Is it because the affiliates are unaware that they need to disclose, or is it because they unaware of the FTC's specific guidelines? Future work could examine this by the means of surveys and detailed interviews with affiliates.

Our work further highlights why having disclosures in affiliate content is important: affiliate content tends to have higher user engagement metrics, which means they are likely to be picked up by generic recommendations algorithms, and shown to users via search or otherwise.

\subsection{Role of Stakeholders in Enabling Disclosures}

\subsubsection{Role of Social Media Platforms}

Along with affiliates, social media platforms play a critical role in shaping the disclosures. Affiliates' disclosures, generally speaking, are limited by the character space available to them. For instance, the description length can be as long as 5000 characters on YouTube, but on Pinterest it is capped to 500 characters. Similarly, tweets can only be as long as 280 characters on Twitter. Therefore, social media platforms can help design their interfaces to make it easier for affiliates to disclose without crowding their promotion text. For instance, Instagram recently added an option for sponsored content to be disclosed by using the ``Paid partnership'' tool, which enables disclosures outside of the traditional image description \cite{instagram}. Similarly, YouTube also added the ability for create a \emph{Contains Product Placement} overlay to their videos \cite{youtube}.

Such disclosure tools are a step in the right direction, however it is unlikely that such blanket disclosures will cover all marketing strategies. Future work could investigate what kind of affordances should be designed into social media platforms to enable affiliates to disclose clearly, and such that users fail to miss the disclosures.

\subsubsection{Role of Affiliate Marketing Companies}

Affiliates could also be held accountable to better disclosure practices by the affiliate marketing companies they sign up for. We examined the affiliate marketing terms and conditions, where publicly available, of eight of the most prevalent affiliate marketing companies from our dataset: Amazon, AliExpress, Commission Junction, Rakuten Marketing, Impact Radius, RewardStyle, ShopStyle and ShareASale. We could not find any publicly available terms and conditions from Impact Radius or RewardStyle. Only Amazon\footnote{https://affiliate-program.amazon.com/help/operating/agreement} and ShopStyle\footnote{https://www.shopstylecollective.com/terms} explicitly referenced the FTC's guidelines in their affiliate terms. While we did not find any references in the Rakuten Marketing and ShareASale affiliate terms, we noted both companies blogging about the guidelines on their company blogs\footnote{http://blog.shareasale.com/2017/09/08/ftc-updates-and-faq-s/}\textsuperscript{,}\footnote{https://blog.marketing.rakuten.com/topic/ftc-disclosure-guidelines}. However, its unclear how many other programs follow this practice, what standards they hold their affiliates accountable to, and whether they explicitly point their affiliates to the FTC's guidelines. Future work could also examine ways in which affiliate marketing companies can reward or penalize those affiliates who fail to adequately disclose.

\subsubsection{Role of Web Browsers}

Finally, Web browsers could help increase transparency into affiliate marketing advertising practices by means of in-built support or through add-ons and extensions. Such tools could function like current Ad-blockers, but rather than blocking advertisements, they could either highlight when a piece of content should contain disclosures, or the disclosures when they are present. Machine learning based approaches naturally lend themselves to developing such tools as various models could be trained on large datasets of social media content.

\subsection{Future Work}

In future work, we hope to take three specific directions. First, we hope to validate the FTC's guidelines on affiliate marketing, and verify that the suggestions they put forward are actually effective in practice. That is, do end-users interpret the disclosures expected by the FTC correctly, and if not, how can we better improve them? Second, our findings show that the three categories of affiliate disclosures often use wordings that exhibit a common pattern. We hope to build a tool---such as a browser extension---that uses machine learning to detect these patterns automatically and highlights them to end-users. Such a tool will be able to help end-users better identify the disclosures that are often buried in the content they view. Third, we hope to examine how the disclosures we discovered align with the disclosure guidelines put forward by agencies in other countries such as the Advertising Standards Authority (ASA) in the United Kingdom.

\section{Limitations}
Our study has two primary limitations. First, we only examined for affiliate marketing disclosures in the description boxes of the YouTube videos and Pinterest pins. Affiliates may also have disclosed their affiliate relationships during the course of the YouTube video, or on the image containing the Pinterest pin. In an initial analysis of 20 randomly selected affiliate videos and pins, we found no disclosures in these locations, and instead focused our attention to the descriptions. In future work, we will expand our analysis to cover both these cases more thoroughly. Second, our method to discover affiliate links should be considered as a lower bound on the number of affiliate marketing programs, since we may have missed the less prevalent companies owing to our frequency analysis. We also did not consider those affiliate programs that use coupon codes to track sales. In future work, we will expand our analysis to encompass a greater set of affiliate programs.

\section{Conclusion}
In this paper, we analyzed affiliate marketing based disclosures on two social media platforms: YouTube and Pinterest. We found that disclosures on these platforms fall into three categories, and the overall rate of disclosure is low. Our findings provide a starting point for several policy based discussions.

\section{Acknowledgements}
Our research was funded by Princeton University. Any opinions, findings and conclusions or recommendations expressed in this material are those of the authors and do not necessarily reflect the views of Princeton University.





%

\bibliographystyle{IEEEtran}
\bibliography{sample}

\section*{Appendix}
\begin{table*}[ht]
  \centering
  \caption{List of Affiliate Marketing Companies Discovered by Our Analysis. Count Indicates the Number of times the URL Pattern Appeared When We Resolved the Retrieved URLs.}
\label{fig:affiliate}
\begin{tabular}{l l l r r}
\toprule
\textbf{Company Name} & \textbf{Domain} & \textbf{URL Pattern} & \textbf{YouTube Count} & \textbf{Pinterest Count}\\

\midrule
\multirow{2}{*}{admitad}  & \multirow{2}{*}{admitad} & https://ad.admitad.com/g/\ldots & \multirow{2}{*}{245}  & \multirow{2}{*}{1} \\
&  & https://ad.admitad.com/goto/\ldots & &  \\[0.15cm]

affiliaXe  & affiliaxe & http://performance.affiliaxe.com/\ldots\&aff\_id=\ldots & 151 & 0 \\[0.15cm]

AliExpress & aliexpress & https://s.aliexpress.com/\ldots\&af=\ldots & 2167 & 785 \\[0.15cm]

Amazon & amazon & http://www.amazon.(com,de,fr,in,it)/\ldots\&tag=\ldots & 7308 & 7368 \\[0.15cm]

Apple & apple & https://itunes.apple.com/\ldots\&at=\ldots & 669 & 61 \\[0.15cm]

Audiobooks & audiobooks & https://affiliates.audiobooks.com/\ldots\&a\_aid=\ldots\&a\_bid=\ldots & 129 & 0 \\[0.15cm]

AvantLink & avantlink & http://www.avantlink.com/\ldots\&pw=\ldots & 34 & 12 \\[0.15cm]

Avangate & avangate & https://secure.avangate.com/\ldots\&AFFILIATE=\ldots & 12 & 0 \\[0.15cm]

\multirow{3}{*}{Awin} & awin1 & http://www.awin1.com/\ldots\&awinaffid=\ldots & \multirow{3}{*}{129} & \multirow{3}{*}{211} \\
& zanox & http://ad.zanox.com/ppc/?\ldots &  & \\
& zenaps & http://www.zenaps.com/rclick.php?\ldots &  & \\[0.15cm]

Banggood & banggood & http://www.banggood.com/\ldots\&p=\ldots & 88 & 13\\[0.15cm]

Book Depository & bookdepository & https://www.bookdepository.com/\ldots\&a\_aid=\ldots & 103 & 0\\[0.15cm]

Booking.com & booking & https://www.booking.com/\ldots\&aid=\ldots & 257 & 7\\[0.15cm]

Clickbank & clickbank & http://\ldots .hop.clickbank.net/\ldots & 678 & 262 \\[0.15cm]

\multirow{8}{*}{CJ Affiliate}  & anrdoezrs & http://www.anrdoezrs.net/click-[0-9]+-[0-9]+\ldots & \multirow{8}{*}{341} & \multirow{8}{*}{2413} \\
 & dotomi & http://cj.dotomi.com/\ldots &  &  \\
 & dpbolvw & http://www.dpbolvw.net/click-[0-9]+-[0-9]+\ldots &  &  \\
 & emjcd & http://www.emjcd.com/\ldots &  &  \\
 & jdoqocy & http://www.jdoqocy.com/click-[0-9]+-[0-9]+\ldots &  &  \\
 & kqzyfj & http://www.kqzyfj.com/click-[0-9]+-[0-9]+\ldots &  &  \\
 & qksrv & http://qksrv.net/\ldots &  &  \\
 & tkqlhce & http://www.tkqlhce.com/click-[0-9]+-[0-9]+\ldots &  &  \\[0.15cm]

Ebay & ebay & http://rover.ebay.com\ldots\&campid=\ldots & 99 & 1963\\[0.15cm]

\multirow{6}{*}{Envato}  & audiojungle & https://audiojungle.net/\ldots\&ref=\ldots & 108 & 0 \\
& codecanyon & https://codecanyon.net/\ldots\&ref=\ldots & 14 & 76 \\
& envato & https://marketplace.envato.com/\ldots\&ref=\ldots & 175 & 262 \\
& graphicriver & https://graphicriver.net/\ldots\&ref=\ldots & 15 & 1465 \\
& themeforest & https://themeforest.net/\ldots\&ref=\ldots & 19 & 200 \\
& videohive & https://videohive.net/\ldots\&ref=\ldots & 578 & 33 \\[0.15cm]

\multirow{2}{*}{e-Commerce Partners Network} & \multirow{2}{*}{buyeasy} & http://buyeasy.by/cashback/\ldots & \multirow{2}{*}{741} & \multirow{2}{*}{7} \\
&  & http://buyeasy.by/redirect/\ldots & & \\[0.15cm]

Flipkart & flipkart & https://www.flipkart.com/\ldots\&affid=\ldots & 81 & 20 \\[0.15cm]

GT Omega Racing & gtomegaracing & http://www.gtomegaracing.com/\ldots\&tracking=\ldots & 56 & 0 \\[0.15cm]

Hotellook & hotellook & https://search.hotellook.com/\ldots\&marker=\ldots & 165 & 5 \\[0.15cm]

Hotmart & hotmart & https://www.hotmart.net.br/\ldots\&a=\ldots & 211 & 8 \\[0.15cm]

\multirow{2}{*}{Impact Radius} & 7eer & http://\ldots.7eer.net/c/[0-9]+/[0-9]+/[0-9]+\ldots & \multirow{2}{*}{180} & \multirow{2}{*}{529} \\
& evyy & http://\ldots.evyy.net/c/[0-9]+/[0-9]+/[0-9]+\ldots & & \\[0.15cm]

KontrolFreek & kontrolfreek & https://www.kontrolfreek.com/\ldots\&a\_aid=\ldots & 117 & 0 \\[0.15cm]

Makeup Geek & makeupgeek & http://www.makeupgeek.com/\ldots\&acc=\ldots & 57 & 0 \\[0.15cm]

\multirow{6}{*}{Pepperjam Network} & gopjn & http://www.gopjn.com/t/[0-9]-[0-9]+-[0-9]+-[0-9]+\ldots & \multirow{6}{*}{2} & \multirow{6}{*}{79} \\
& pjatr & http://www.pjatr.com/t/[0-9]-[0-9]+-[0-9]+-[0-9]+\ldots &  &  \\
 & pjtra & http://www.pjtra.com/t/[0-9]-[0-9]+-[0-9]+-[0-9]+\ldots &  &  \\
 & pntra & http://www.pntra.com/t/[0-9]-[0-9]+-[0-9]+-[0-9]+\ldots &  &  \\
 & pntrac & http://www.pntrac.com/t/[0-9]-[0-9]+-[0-9]+-[0-9]+\ldots &  &  \\
 & pntrs & http://www.pntrs.com/t/[0-9]-[0-9]+-[0-9]+-[0-9]+\ldots &  &  \\[0.15cm]

Rakuten Marketing & linksynergy & http://click.linksynergy.com/\ldots\&id=\ldots & 189 & 1877 \\[0.15cm]

Skimlinks & redirectingat & http://go.redirectingat.com/\ldots\&id=\ldots & 43 & 155 \\[0.15cm]

Smartex & olymptrade & https://olymptrade.com/\ldots\&affiliate\_id=\ldots & 65 & 0 \\[0.15cm]

RewardStyle & rstyle & http://rstyle.me/\ldots & 402 & 2711 \\[0.15cm]

ShopStyle & shopstyle & http://shopstyle.it/\ldots & 111 & 9239 \\[0.15cm]

\multirow{3}{*}{ShareASale} & \multirow{3}{*}{shareasale} & http://www.shareasale.com/r.cfm\ldots & \multirow{3}{*}{199} & \multirow{3}{*}{616} \\
 &  & http://www.shareasale.com/m-pr.cfm\ldots &  &  \\
 &  & http://www.shareasale.com/u.cfm\ldots &  &  \\[0.15cm]

Studybay & apessay & https://apessay.com/\ldots\&rid=\ldots & 141 & 0 \\[0.15cm]

Zaful & zaful & http://zaful.com/\ldots\&lkid=\ldots & 32 & 786 \\[0.15cm]

\bottomrule
\end{tabular}
\end{table*}

\end{document}